# Unsupervised Ensembling of Multiple Software Sensors with Phase Synchronization: A Robust Approach For Electrocardiogram-derived Respiration

**Jacob McErlean[1], John Malik[1], Yu-Ting Lin[2,3], Ronen Talmon[4], Hau-Tieng Wu[1,5†]**

[1] Department of Mathematics, Duke University, Durham, North Carolina, USA

[2] School of Medicine, National Yang Ming Chiao Tung University, Taipei, Taiwan

[3] Department of Anesthesiology, Taipei Veterans General Hospital, Taipei, Taiwan

[4] Faculty of Electrical and Computer Engineering, Technion - Israel Institute of Technology, Haifa, Israel

[5] Department of Statistical Science, Duke University, Durham, North Carolina, USA

Email: †hauwu@math.duke.edu

Present address: Department of Mathematics, Duke University, Science Dr., Durham, NC, 27708, USA

## ABSTRACT

***Objective:*** We aimed to fuse the outputs of different electrocardiogram-derived respiration (EDR) algorithms to create one EDR signal that is of higher quality.

***Methods:*** We viewed each EDR algorithm as a software sensor that recorded breathing activity from a different vantage point, identified high-quality software sensors based on the respiratory signal quality index, aligned the highest-quality EDRs with a phase synchronization technique based on the graph connection Laplacian, and finally fused those aligned, high-quality EDRs. We refer to the output as the sync-ensembled EDR signal. The proposed algorithm was evaluated on two large-scale databases of whole-night polysomnograms. We evaluated the performance of the proposed algorithm using three respiratory signals recorded from different hardware sensors, and compared it with other existing EDR algorithms. A sensitivity analysis was carried out for a total of five cases: fusion by taking the mean of EDR signals, and the four cases of EDR signal alignment without and with synchronization and without and with signal quality selection.

***Results:*** The sync-ensembled EDR algorithm outperforms existing EDR algorithms when evaluated by the synchronized correlation ($\gamma$-score), optimal transport (OT) distance, and estimated average respiratory rate (EARR) score, all with statistical significance. The sensitivity analysis shows that the signal quality selection and EDR signal alignment are both critical for the performance, both with statistical significance.

***Conclusion:*** The sync-ensembled EDR provides robust respiratory information from electrocardiogram.

***Significance:*** Phase synchronization is not only theoretically rigorous but also practical to design a robust EDR.

## 2 INTRODUCTION

Breathing is the physical process that facilitates the transfer of gases in and out of cells, while respiration is the cellular process of converting glucose to energy. Below, we follow the broadly accepted practice and view the clinical terms *respiratory signal* and *breathing signal* as any signal containing information related to the breathing process. Respiratory signals allow clinicians to measure quantities such as tidal volume[1], breathing rate, and the variability of the breathing pattern [2]. Breathing rate can be assessed by counting the number of breaths a patient takes in a one-minute period [3, 4]. However, effectively assessing breathing pattern variability requires more signal processing [5, 6]. These quantities can be used to monitor respiratory health or to predict respiratory failure [7].

Spirometry produces the spirogram, the gold standard among respiratory signals. However, spirometry is uncomfortable and not ideal for long-term monitoring [8]. Beyond the spirometer, a variety of hardware and software sensors for measuring breathing activity have been proposed [9], including pressure sensors in the nasal cannula, piezosensors measuring thoracic and abdominal movement, chemical capnography sensors for tidal CO2 measurement and even *software sensors* indirect from electrocardiographic (ECG) or photoplethymographic (PPG) sensors. Compared to the spirogram, the information obtained from these sensors might be less reliable. In this paper, we focus on a class of software sensor that provides indirect respiratory signals that can be derived by software from the single-channel ECG. This capability is based on the fundamental physiological fact that when the patient inhales (resp. exhales), transthoracic electrical impedance increases (resp. decreases). As the patient breathes in and out, the shape of the ECG waveform changes accordingly. As a result of this capability, ECG sensors are easy-to-use, mobile, and non-invasive tools for measuring breathing activity.

Deriving a respiratory signal from multiple- or single-channel ECG has been attempted using several methods [10-12]. Traditionally, an EDR signal is obtained by interpolating the time series of RS amplitudes as follows:

$$EDR\left(\frac{r_i}{f_s}\right) := E(r_i) - E(s_i), \qquad (1)$$

where $E \in R^n$ is the discrete-time ECG signal having length $n$ and a sampling rate of $f_s$ Hz, $r_i$ is the sample index of the $i$-th R peak, and $s_i$ is the sample index of the subsequent S peak. See Figure 1 for an illustration. EDR algorithms based on the upward and downward slopes of the QRS complex [45] have been proposed, as well as on the R-wave angle formed by these slopes [46, 47], and a recent study [56] showed that these methods outperforms traditional methods. Another method involves estimating the area of each QRS complex [12] followed by interpolation. Principal component analysis (PCA) [13, 14], independent component analysis (ICA) [15], and kernel principal component analysis (kPCA) [16] have also been used to obtain an EDR signal. Recently, an EDR algorithm based on diffusion maps (DM) [17] was proposed in [48].

While there have been several algorithms, each has its own limitations. For example, if the R peak detection is not accurate due to any reason, if the R or S peak amplitude is contaminated by noise, or when there exist arrhythmic beats like premature ventricular contractions or atrial fibrillation [18, 47], the EDR signal based on RS amplitudes might be distorted. Similarly, QRS slope based algorithms [45,46,47] depend on QRS complex morphology, which is impacted by ECG signal quality. The PCA, kPCA or DM-approaches might be sensitive to noise, which is of high dimensional nature. Hence, it is possible that one EDR algorithm outperforms the other in certain scenarios, while the performance dynamics may reverse in different situations, and identifying a superior algorithm for all situations is challenging. For example, one comparison article advocates measuring the difference in potential between the R wave and the S wave (the RS amplitude) [19], while another advocates measuring the area of each QRS (ventricular depolarization) complex [20]. Ideally, a high-quality EDR signal must allow the clinician to accurately measure clinical indicators, including breathing rate, breathing pattern variability, or even tidal volume, etc.

Since each EDR algorithm has its own pros and cons, and since finding a superior one is challenging, an intuitive idea is combining different algorithms so that the strength of each algorithm can be preserved while limitations can be mitigated. The proposed algorithm in this paper is inspired by the concept of ensemble learning (or sensor fusion) in machine learning [21]. The main contribution of this work is a software methodology for fusing multiple simultaneous estimates of the breathing activity that were derived from an ECG sensor. We output a new EDR signal (coined the *sync-ensembled EDR signal*) that is of higher quality than each input EDR signal in the sense of capturing the breathing oscillation. Specifically, we view each EDR algorithm as a *software sensor* that recorded breathing activity from a different vantage point, identify high-quality software sensors based on the respiratory signal quality index, align those high-quality EDRs with a *synchronization* technique motivated by the graph connection Laplacian (GCL), and fuse those aligned, high-quality EDRs.

We remark that in the machine learning literature, sensor fusion approaches may be classified as *early*,

*halfway*, or *late* depending on when the fusion takes place. We view individual EDR algorithms as software sensors that *sense* breathing activity from an ECG recording. Under this paradigm, an early sensor fusion method would create *one* EDR signal from directly integrating information from multiple ECG leads. Conversely, a late sensor fusion algorithm would be one that computes the mean of multiple estimates for clinical features, e.g., breathing rate. Our approach may be considered a halfway approach because it occurs after many individual EDR estimates are created but before indicators such as breathing rate and breathing pattern variability are inferred.

## 3 Methods

We provide a detailed description of our proposed sync-ensembled EDR algorithm that extracts breathing activity from an ECG signal and the rationale behind it. The flowchart of the algorithm is shown in Figure 2. The algorithm is divided into three steps, which we detail below. The MATLAB code and partial database are available on Github (https://github.com/jacobmcerlean/EDRsynchronization).

### *3.1 Proposed algorithm*

### *3.2 Step 1: Pre-processing*

The raw ECG signal is first upsampled to a sampling rate of 1000 Hz [22]. Measurement noise and powerline interference in the ECG signal are suppressed by applying a bi-directional, third-order Butterworth lowpass filter with a cutoff frequency of 40 Hz. Baseline wandering is estimated and subsequently removed by applying a median filter with a window size of 200 ms to each channel of the ECG signal. This step also moderately suppresses the P and T waves. We detect the QRS complexes in the ECG signal by applying a high-accuracy QRS detector [23]. Denote the pre-processed ECG signal as a vector $E \in R^n$, where $n = f_s \times T$ is the number of samples, $f_s = 1000$ Hz is the sampling rate of the signal, and $T$ is the duration of the recording in seconds. Suppose there are $N$ detected R peaks in the ECG signal. Let $r_i$ denote the location in samples of the $i$th R peak in the ECG (Or the $i$th S peak in the ECG if the cardiac axis is deviated). To analyze the traditional EDR signal, the S peak following each R peak is determined by

$$s_i = argmin_{r_i+1 \leq t \leq r_i+60} E(t), \qquad (2)$$

where the window length of 60 samples, corresponding to a 60 ms duration, is chosen according to normal electrophysiology. Similarly, the Q peak preceding each R peak is determined by

$$q_i = argmin_{r_i-30 \leq t \leq r_i-1} E(t), \qquad (3)$$

identified in the 30 ms window preceding the R peak.

By defining a window around each R peak that extends 30 samples (30 ms) to the left and 60 samples (60 ms) to the right, we are able to capture the entire QRS complex, recalling that the P and T waves have been suppressed. For each QRS estimate, we collect the segments containing QRS complexes and stack them as rows in a matrix:

$$X(i, j+31) \coloneqq E(r_i + j), \quad -30 \leq j \leq 60, \tag{4}$$

where the resulting matrix is $X \in R^{N \times p}$, where $p = 91$. Note that the ordering of the rows of $X$ encodes temporal information.

### *3.3 Step 2: Estimating the respiratory signal*

We build a variety of estimates of the respiratory signal; these estimates, resulting from several techniques, are viewed as *software sensors* for the breathing activity. We considered the traditional approach based on RS amplitudes (1), the downward slope of the QRS complex (DW), the R-wave angle (RA), the geometric approach (PCA), and the manifold learning approach (DM), to obtain a pool of EDR signals. All estimates were rendered at a sampling rate of 10 Hz. Subsequently, the signal quality of these estimated respiratory signals from various software sensors was evaluated, and high-quality EDR signals were aligned and fused to obtain a new EDR signal, the *ensembled EDR*. Below, we describe these estimates in detail.

#### *3.3.1 Traditional approach*

The first estimate for the respiratory signal is the traditional EDR signal. Set

$$R^{trad}\left(\frac{r_i}{f_s}\right) = E(r_i) - E(s_i) \tag{5}$$

where $i = 1, \ldots, N$. Then apply the piecewise cubic spline interpolation to the assigned $N$ points $\left\{R^{trad}\left(\frac{r_i}{f_s}\right)\right\}_{i=1}^{N}$, viewed as a set of nonuniform sampling points, and sample the resulting signal at a sampling rate of 10 Hz. This yields an interpolated vector $R^{trad} \in R^{\lfloor 10T \rfloor}$. This interpolation is applied to all the subsequent EDR estimates, and thus, for brevity, is omitted.

### *3.3.2 Downward slope approach* [45]

Let $R_i$ and $S_i$ denote the indices for R- and S-peaks, respectively. On the interval $R_i$ to $S_i$, the steepest downward slope of the QRS complex is identified at time $s_i$, and a straight-line with slope $d_i$ is fit to the signal within an 8 ms window centered at $s_i$. Denote the line as $D_i$. Then, the EDR estimate at time $s_i$ is set to $d_i$. Then apply the piecewise cubic spline interpolation to the assigned $N$ points $\left\{\left(\frac{s_i}{f_s}, -d_i\right)\right\}_{i=1}^{N}$, viewed as a set of nonuniform sampling points, and sample the resulting signal at a sampling rate of 10 Hz.

### *3.3.3 R-wave angle approach* [46, 47]

Identify the time $s_i$ with the steepest slope of the QRS complex between the $i$th Q- and R-peaks, and fit a line within an 8 ms window centered at $s_i$, with the upward slope $u_i$. Denote the line as $U_i$. With the fit line $D_i$ in 3.3.2, the R-wave angle between $D_i$ and $U_i$ is taken to be

$$RW_i = arctan(|u_i - d_i|/(1 + u_i * d_i)) . \qquad (6)$$

Then apply the piecewise cubic spline interpolation to the assigned $N$ points $\left\{\left(\frac{s_i}{f_s}, RW_i\right)\right\}_{i=1}^{N}$, viewed as a set of nonuniform sampling points, and sample the resulting signal at a sampling rate of 10 Hz.

### *3.3.4 Geometric approach*

The next two estimates are obtained via PCA. Let $p_1, \ldots, p_5 \in R^p$ be the top five principal components of the centralized matrix $X$ (that is, the top five eigenvectors of the corresponding $p \times p$ covariance matrix). We then set

$$R_j^{PCA}\left(\frac{r_i}{f_s}\right) = \sum_{k=1}^{p} \tilde{X}(i,k)p_j(k),\ j = 1, \ldots, 5. \qquad (7)$$

where $\tilde{X}$ is the decentralized $X$. Details about this method can be found in [13, 14]. We mention that the number five is chosen in an ad hoc way. A different number of principal components can be considered.

### *3.3.5 Manifold learning approach*

Next, we apply a nonlinear manifold learning method called the DM algorithm [17, 24], which can be viewed as a generalization of the kernel PCA approach [16]. We build a self-tuning affinity matrix $W \in R^{N\times N}$ by setting

$$W(i,j) = \frac{1}{2}exp\left[-\sum_{k=1}^{p} \frac{(X(i,k)-X(j,k))^2}{\sigma_i}\right] + \frac{1}{2}exp\left[-\sum_{k=1}^{p} \frac{(X(i,k)-X(j,k))^2}{\sigma_j}\right], \qquad (8)$$

where $\sigma_i$ is the median of the squared Euclidean distances from the $i$-th row of $X$ to all other rows. We then perform $\alpha$-normalization with $\alpha = 1$ to mitigate any non-uniform densities (see [17] for details). We calculate

$$W_1 = D^{-1} W D^{-1}, \qquad (9)$$

where $D \in R^{N \times N}$ is a diagonal matrix satisfying $D(i,i) = \sum_{j=1}^{N} W(i,j)$. We then calculate the isotropic diffusion kernel by setting

$$P = D_1^{-\frac{1}{2}} W_1 D_1^{-\frac{1}{2}}, \qquad (10)$$

where $D_1 \in R^{N \times N}$ is a diagonal matrix satisfying $D_1(i,i) = \sum_{j=1}^{N} W_1(i,j)$. We obtain the top five non-trivial eigenvectors $\phi_1, \ldots, \phi_5 \in R^N$ of $P$, and we set $\varphi_j \in R^N$ to be the vector

$$\varphi_j = D^{-\frac{1}{2}} \phi_j. \qquad (11)$$

Finally, we set

$$R_j^{DM}\left(\frac{r_i}{f_s}\right) = \varphi_j(i),\ j = 1, \ldots, 5. \qquad (12)$$

Again, we mention that a different number of eigenvectors can be considered.

### *3.4 Step 3: Screening signals for ensemble using RQI*

From the ECG signal input, thirteen EDR estimates are derived, including one signal from the traditional EDR estimate, the downward slope estimate, the R-wave angle estimate, five from the PCA estimate, and five from the diffusion map estimate. However, not all of them are of high quality. Clearly, the quality of individual EDR estimate would impact the quality of the final ensembled signal. Therefore, we consider excluding EDR estimates of poor signal quality from the next ensemble step. Several methods to assess respiratory signal quality exist, and the one we consider is to analyze the strength of the signal's power spectrum peak within the typical breathing rate range [25]. We consider this RQI to balance the speed and accuracy. The RQI of each EDR estimate is estimated in the following steps. First, each 2-minute EDR signal is linearly de-trended. Then, the signal is passed through a third-order Butterworth bandpass filter between 0.1 and 0.75 Hz. Afterwards, the signal's power spectrum, denoted as a vector $P(i)$, is computed with a fast Fourier Transform. The total respiratory energy (TRE) is set to be the sum of power within the range 0.1-0.75 Hz. The choice of lower and upper frequency limits for this range are taken directly from [25]. Then, the index of the peak of the power spectrum is identified and denoted by the index *m*. The maximum peak energy (MPE) is set to be

$$MPE = P(m-1) + P(m) + P(m+1). \qquad (13)$$

Consequently, the RQI of the signal is defined

$$RQI = \frac{MPE}{TRE}. \qquad (14)$$

By construction, the RQI metric occupies a range of 0 to 1, with a higher metric indicating a signal with a stronger peak within the typical breathing rate range, and therefore more likely to encode higher quality breathing information. A threshold of 0.2 was chosen for the EDR estimates, so that of the 13 EDR estimates, only those with an RQI of at least 0.2 were designated as desirable for the ensemble. If at least five EDR estimates had RQIs above the threshold, then only the EDR estimates above the threshold were included in the ensemble, independent of their algorithms of origin. If fewer than three EDR estimates had RQIs above the threshold, then the ensembled signal was formed by combining all EDR estimates above the threshold together with the individual signal of highest RQI from any of the five algorithms not already contributing to the ensemble. For the purpose of demonstrating the proposed algorithm, this procedure guarantees that the ensembled signals are genuinely ensembles, as the ensembled signals are always formed from at least three individual EDR signals. The choice of threshold is chosen to exclude bad quality segments and the result is confirmed by visual inspection, and we do not tune the threshold to optimize the performance of the ensembled signal.

### *3.5 Step 4: Synchronizing the estimated respiratory signals*

Prior to combining the $L \geq 1$ remaining high quality EDR signals, we normalize them. For the $k$-th EDR estimate $R_k \in R^{\lfloor 10T \rfloor}$, set

$$\tilde{R}_k(i) = \frac{R_k(i) - \sum_{k=-49}^{50} R_k(i+k)}{\sqrt{\sum_{l=-49}^{50}\left(R_k(l) - \sum_{m=-49}^{50} R_k(l+m)\right)^2}}, \qquad (15)$$

where $i = 1, \dots, \lfloor 10T \rfloor$ and we set $R_k(i) = 0$ when $i < 1$ or $i > \lfloor 10T \rfloor$. This procedure corresponds to detrending and locally $z$-scoring the signal.

Due to the inevitable global phase deviation and erroneous breathing cycle recovery in each EDR estimate, the main novelty of this paper is applying the graph connection Laplacian [26] to adjust the global phase deviation and align all EDR estimates. We first estimate the pairwise phase deviation. For any two EDR estimates $\tilde{R}_i$ and $\tilde{R}_j$, we compute

$$c_{i,j} := \frac{\langle H(\tilde{R}_i),\ \langle H(\tilde{R}_j)\rangle}{|\langle H(\tilde{R}_i), H(\tilde{R}_j)\rangle|}\,, \qquad (16)$$

where $H$ is defined as $H(\tilde{R}_i) = \tilde{R}_i + i \times Hilbert(\tilde{R}_i)$, with $Hilbert$ the Hilbert transform, and $\langle\,,\rangle$ is the inner product of two complex signals. The entry $c_{i,j}$ estimates the global phase shift of $\tilde{R}_i$ and $\tilde{R}_j$. Then, suppose we have $L$ EDR estimators. Construct a Hermitian function $C$ of size $L \times L$ so that the $(i,j)$th entry of $C$ is $c_{i,j}$. Note that this $C$ matrix is directly related to the GCL discussed in [26, 27]. Find the eigenvector of $C$, denoted as $u$, that is associated with the largest eigenvalue. The global phase deviation of the $i$th EDR estimate is then corrected by

$$\bar{R}_i := Re[H(\tilde{R}_i) \times \overline{u(i)}]\,, \qquad (17)$$

where $Re$ means taking the real part.

### 3.6 Step 5: Obtain the final EDR signal by ensembling

We proceed with ensembling the estimates using a method that can be extended to include any type of EDR signal. Consider the initial matrix $M^{(0)} \in R^{\lfloor 10T\rfloor \times L}$, whose columns are the remaining high quality EDR estimates that have been phase adjusted by the above RQI. Then, we compute the time-lagged representations $M^{(t)} \in R^{(\lfloor 10T\rfloor - t)\times L}$, given by

$$M^{(t)}(i,j) := M^{(0)}(i+t,j) \qquad (18)$$

for all $i = 1, \dots, \lfloor 10T\rfloor - t$, $j = 1, \dots, L$, and $t = 0, \dots, 9$. We define a matrix $B$ with $10L$ columns, whose rows are given by

$$row_i B := \left[row_i M^{(0)} \cdots row_i M^{(8)}\, row_i M^{(9)}\right] \qquad (19)$$

for all $i = 1, \dots, \lfloor 10T\rfloor - 9$. The proposed estimate for the respiratory signal, the ensembled EDR, is the top left-singular vector of $B$, denoted as $U \in R^{\lfloor 10T\rfloor - 9}$. Note that the first entry represents the estimated respiratory volume at time 1.0 seconds; to compensate for this lag, we add 9 null values at the beginning of the signal $U$, since the effective sampling rate is 10 Hz. We thus obtain the proposed EDR signal.

### 3.7 Rationale of the proposed synchronization approach

The rationale underlying the alignment by the GCL-motivated synchronization and singular value decomposition (SVD) for our sync-ensemble EDR algorithm is as follows. The main observation is that even being selected by RQI, different remaining EDR algorithms capture the breathing activity with different qualities. We could model each EDR signal as a respiratory signal contaminated by undesired noise, where the noise type and level vary from one EDR to another. Moreover, due to the nature of different algorithms, these EDR signals, even if the quality is high, might have phase deviations, and the phase deviation can be large.

To correct the global phase deviation when the EDR signals are noisy, the GCL-motivated synchronization scheme is applied. The underlying philosophy is based on a simple model. Suppose all EDR estimates are perfect without any errors, except a global phase deviation. Denote the global phase deviation of the $j$th EDR estimates associated with the true but inaccessible breathing activity, is $\varphi_j \in [-\pi, \pi]$. This phase deviation can be represented as a complex number $e^{i\varphi_j}$. Thus, the true global phase deviation of 13 perfect EDR estimates can be represented as a 13-dimensional complex vector $v$, whose $j$th entry is $e^{i\varphi_j}$. Note that the global phase relationship between the $j$th and $k$th EDR estimates is $\varphi_j - \varphi_k$, which can be represented as a complex number $e^{i(\varphi_j - \varphi_k)}$. We can see that $e^{i(\varphi_j - \varphi_k)}$ is nothing but the $(j, k)$th entry of the rank-1 matrix $vv^*$, where $v^*$ is the conjugate transpose of $v$. This suggests that if we can accurately acquire $e^{i(\varphi_j - \varphi_k)}$, then the global phase deviation problem can be resolved.

However, due to the inevitable errors induced by the noise, arrhythmia, and other artifacts, we cannot accurately obtain $e^{i(\varphi_j - \varphi_k)}$ via (16). What the best we can obtain via (16) is a noisy version of a $L \times L$ submatrix of $C$, where $L$ is the number of high quality EDR signals selected by RQI. It is natural to ask how accurately we can recover the global phase deviation for those $L$ EDR signals. The stability property of this kind of matrix has recently been established [27, 28], which states that the top eigenvector of $C$ provides reliable global phase information under such noisy situations. Thus, via (17), the issue of global phase deviation among different EDR estimates is resolved. We refer readers with interest in this topic to [26-29] and the cited papers therein for more details.

Even after the correction by the synchronization, there might still be small phase deviation remaining, despite the stability of $C$. To take care of this small phase deviation, the matrix $B$ is constructed, which includes possible phase deviations of each EDR. Thus, the most phase aligned respiratory signal constitutes the dominant left singular component in $B$. Geometrically, $U$ represents the direction (in the high-dimensional space $R^{\lfloor 10T-9 \rfloor}$) at which the power of all of the EDR estimates is maximized. In other words, $U$ represents the common information between all of the EDR estimates. According to the SVD, $U$ is the maximum argument of $\max_{\|U\|=1} \|U^\top B\| = \max_{\|U\|=1} \sqrt{\sum_{i=1}^{10Q} \langle U, col_i B \rangle^2}$. As $\langle U, col_i B \rangle$ can be viewed as the correlation between $U$ and $col_i B$, we can intuit that $U$ is the signal that is most correlated with the columns of $B$. This model justifies the application of the SVD. More explanations of the proposed algorithm can be found in the Discussion section.

*3.8 Material*

The ECG signals featured in our experiment are from two databases.

The first is a set of 50 standard overnight polysomnograms which were recorded to confirm the presence of sleep apnea syndrome in subjects suspected of sleep apnea at the sleep center in Chang Gung Memorial Hospital (CGMH), Linkou, Taoyuan, Taiwan. The Institutional Review Board of CGMH approved the study protocol (No. 101-4968A3), and the enrolled subjects provided written informed consent. This database is part of the Taiwan Integrated Database for Intelligent Sleep (TIDIS). The research was conducted in accordance with the principles embodied in the Declaration of Helsinki and in accordance with local statutory requirements. All recordings were acquired on the Alice 5 data acquisition system (Philips Respironics, Murrysville, PA). Of these 50 subjects, there were 26 male patients with the age range 12-72 (46.27 $\pm$15.79) years old, and 24 female patients with the age range 19-67 (43.54 $\pm$ 15.59) years old. All subjects had an apnea-hypopnea index less than 5. We use the first ECG recording, which, before upsampling, was sampled at 200 Hz. To evaluate the effectiveness of our estimate for the respiratory signal, we compare it to the airflow (denoted as **CFLOW)**, the thoracic movement (**THO)**, and the abdominal movement (**ABD**) signals, where **CFLOW** is the patient flow signal recorded from a nasal cannula, **THO** is recorded from a chest band with a piezosensor, and **ABD** is recorded from an abdominal band with a piezosensor. All reference signals are sampled at 10 Hz.

The second database is the Sleep Heart Health Study (SHHS) [49, 50] a multi-center cohort study implemented by the National Heart Lung & Blood Institute to determine the cardiovascular and other consequences of sleep-disordered breathing. The database is formed from a total of 6,441 men and women aged 40 years and older who were enrolled between November 1, 1995 and January 31, 1998 in SHHS Visit 1. From the subjects of SHHS1, the first 250 subjects of the database were selected as a subset for the experiment. Of these 250 subjects, there were 134 male patients with the age range 40-83 (58.66 $\pm$10.90) years old, and 116 female patients with the age range 40-84 (58.66 $\pm$ 12.04) years old. ECG leads were sampled at 125 Hz, and the **CFLOW** (nasal-oral thermocouple, Protec, Woodinville, WA) reference signals were sampled at 10 Hz.

*3.9 Evaluation*

To evaluate the performance of our algorithm and compare it to the performance of other algorithms, we employ the following procedure. For each reference signal (**CFLOW**, **ABD**, or **THO**), the signal was cut into non-overlapping 2-minute segments. In order to evaluate performance of the individual and ensemble EDR algorithms, a high-quality reference signal is required for validation. A segment was screened from the experiment if its RQI was lower than a threshold of 0.6. This ensured that the reference signals tested were of sufficiently high quality to reliably evaluate algorithm performance. Note that the signal quality of the ECG is not taken into consideration. The total numbers of 2-minute segments for the CGMH database **CFLOW, ABD** and **THO** are 803, 246, and 483, respectively. The total number of 2-minute segments for the SHHS database **CFLOW, ABD** and **THO** are 3220, 4832, and 4727, respectively. We build our proposed sync-ensembled EDR on each 2-minute segment. Let $U$ denote the sync-ensembled EDR for one 2-minute segment, and let $V$ denote either the **CFLOW** signal, the **THO** signal, or the **ABD** signal. For each integer lag parameter $\tau$ in the range $-20 \leq \tau \leq 20$, we consider the phase-shifted signal

$$U^{(\tau)}(j) := Re(e^{\pi i \tau/10} U(j)) \qquad (20)$$

and calculate the absolute value of the Pearson correlation coefficient $\rho_\tau$ between $U^{(\tau)}$ and $V$. Our final measure of similarity between $U$ and $V$ is

$$\gamma = \max_{-20 \leq \tau \leq 20} [100 \times \rho_\tau] \qquad (21)$$

The metric $\gamma$ is inherited from a previous work which examined the use of kernel PCA for building an EDR signal [16]. Note that the parameter $\tau$ allows us to account for the unknown global phase discrepancy between the sync-ensembled EDR and the respiratory signal as measured through the various sensors.

Inspired by a previous work [33], we introduce a second evaluation metric to quantify the extent to which our sync-ensembled EDR signal can accurately measure *time-varying breathing rate*. Effectively, it is a comparison between the time-frequency representations (TFRs) of $U^{(\tau^*)}$ and $V$, where $\tau^*$ is maximum argument of (21). We use the de-shape synchrosqueezing transform (dsSST) [34] to calculate both TFRs. The dsSST is designed to remove harmonics in the spectrogram that result from the non-sinusoidal nature of the breathing pattern. Note that these harmonics are not relevant when assessing instantaneous breathing rate. We use a window size of 200 samples, a Gaussian window with bandwidth 0.15, a soft log parameter of 0.03, a hop of 1, and 300 DFT points in the dsSST. The columns of both TFRs are $l^1$-normalized and compared using the optimal transport (OT) distance [35], yielding the metric $\eta$.

To further evaluate the performance of EDR signals, we additionally use a score based on the percentage relative accuracy of the average frequency of the signals compared to the reference. In particular, let $\bar{f}_{i,R}$ be the average frequency of the reference signal, and $\bar{f}_i$ be the average frequency of EDR estimate *i.* Then, the computed score $EARR_i$ based on average frequency for this estimate is the percentage accuracy derived from relative error

$$EARR_i = 100 - 100 \times |\bar{f}_{i,R} - \bar{f}_i| / \bar{f}_{i,R} \quad (22)$$

The average frequency of signals was computed using cycle-detection code from [51], which implements the methods of [52,53].

To determine if one method is better than the other, we apply the one-sided Wilcoxon signed rank test with the significance level set to $p = 0.05$, and we apply the Bonferroni correction to handle the multiple testing issue.

We carry out a sensitivity analysis by considering four different scenarios to validate the necessity of each step in the proposed algorithm. In the first scenario, the ensembled signal $U_1$ was obtained by naively taking the pointwise mean of all 13 EDR estimated signals. The second scenario $U_2$ fused the signals using the time-alignment ensembling approach outlined in Step 5 of the proposed algorithm of the signals without RQI screening. In the third scenario, the GCL was used to phase-align all 13 individual EDR estimates to obtain $U_3$, but we still do not select high quality EDR estimates. The fourth scenario obtains signal $U_4$ without any phase-alignment of signals, but with RQI screening; this mimics the Smart Fusion proposed in [54]. Finally, the full algorithm produces the ensembled signal $U$. We can summarize which algorithm steps the various ensembled signals use and omit: $U_1$ corresponds to algorithm Steps 1-2 only, and then takes the mean of all EDR signal estimates; $U_2$ omits Steps 3 (RQI screening) and 4 (phase-synchronization); $U_3$ omits Step 3 (RQI screening); $U_4$ omits Step 4 (phase-synchronization); and, $U$ is the full algorithm, Steps 1-5.

## 4 Results

The Matlab implementation of the algorithm for results shown in this section can be found online (https://github.com/jacobmcerlean/EDRsynchronization). We begin by illustrating results for visual inspection. In Figure 4, we show a short segment of the EDR derived by the proposed sync-ensemble algorithm and the accompanying **CFLOW**, **THO**, and **ABD** for a visual comparison. We could see that the respiratory cycles could be clearly captured by the proposed algorithm. In Figure 5, we show a comparison of various EDR algorithms and our proposed ensemble algorithm. Although the traditional EDR is selected, we can visualize that its quality is lower than other selected ones. This example shows that the ensemble step in the proposed algorithm helps obtain a higher quality EDR.

To evaluate the performance of our proposed ensemble EDR approach, we compare it with existing EDR algorithms. For each ECG segment input, we obtained 13 EDR estimates: one from the traditional

EDR algorithm, one from the downward slope algorithm (DW), one from the R-wave angle algorithm (RA), five EDR estimates from a dimension-five PCA-based algorithm, and five EDR estimates from a dimension-five DM-based algorithm.

In Table 1 for the CGMH database and Table 3 for the SHHS database, the final EDR derived by various traditional EDR algorithms, the proposed ensemble algorithm is compared to the accompanying **CFLOW**, **THO**, and **ABD** signals in the polysomnogram (only **CFLOW** for the SHHS data). In these tables, the results are shown on the subject level; that is, we show the mean of $\gamma$-indices (and EARR scores) among the subjects ± mean of standard deviations of $\gamma$-indices (and EARR scores) among the subjects and the median of OT distances among the subjects ± median absolute deviation of subject median absolute deviations of OT distances among the subjects. Higher $\gamma$-indices and EARR scores, and lower OT distances, indicate a more accurate recovery of the respiratory signal.

In Table 2 for the CGMH database and Table 4 for the SHHS database, the ensembled signal is compared to the ensembled signal results from the four scenarios for sensitivity analysis, and again evaluated by EARR score, $\gamma$-indices, and OT distances and shown in the same way as that in Tables 1 and 3.

To sum up, the proposed sync-ensembled EDR consistently performed better with statistical significance compared with any single EDR algorithm. The sensitivity analysis shows that both the GCL-based alignment and the signal quality screening are important.

To have a deeper understanding of this result, we show an error plot of the mean $\gamma$-score $\pm$ standard deviation for each of the subjects in the SHHS database whose segments were not screened based on **CFLOW** RQI quality. These are ordered by the mean $\gamma$-score, in Figure 6. We could see that except some cases, most cases perform well with the mean $\gamma$-score higher than 80. For those cases with poor performance, their EDR signals are usually of low quality; that is, few channels are selected by the RQI step. Next, we explore how the RQI plays a role in the proposed algorithm. In Figure 7, for all 2-minute **CFLOW** segments, we compare the number (out of 13) of selected EDR signals above the RQI threshold of 0.2 to the $\gamma$-score of the resulting ensembled signal, and the distribution of the number (out of 13) of selected EDR signals is shown in Figure 8. We can clearly see that the quality of the ensembled EDR increases with the number of selected high-quality EDRs, and we can see an elbow pattern showing up when the number of high quality EDRs is 4. Also, according to Figure 8, for most segments, our proposed algorithm selects 7~10 EDR signals in the final ensemble.

## 5 Discussion

In this work, we demonstrate that a quality-screened and phase-aligned ensemble of multiple EDR signals can lead to a better estimate for the respiratory signal, which we call the sync-ensembled EDR signal. The traditional EDR signal is fused with the EDR signals from DW and RA algorithms, as well as the EDR signals obtained via PCA and via DM, incorporating phase- alignment and signal quality selection, to achieve an estimate for the respiratory signal which exceeds all individual estimates. As was discussed in detail above, the motivation for the phase alignment comes from the commonplace clinical reality of comparing signals with time-varying instantaneous frequencies. The usefulness for RQI screening is inherent to the intuition of the algorithm that seeks to recover information from multiple sources: the ensembled signal's quality depends on the quality of its input EDR signals. In our algorithm, all parameters were chosen *ad hoc* without any optimization or search. In particular, there is freedom to experiment with different bandwidths of the Gaussian kernel used in the DM algorithm, or to use the previously proposed selection criteria [16]. An optimization approach may yield stronger results but also may lead to overfitting at the database level. While this study uses a single ECG-lead, we remark that the approach is independent of the number of ECG-leads, and could potentially be used for a greater number of ECG-leads. Incorporating additional ECG leads offers a direct advantage in addressing the problem of missing leads. When only a single ECG lead is available, no algorithm can cope with missing data for that lead. However, this issue can be effectively managed if at least one other ECG lead is accessible. Finally, we mention that the fusion idea, the phase-alignment and the signal quality screening considered in the proposed ensemble EDR algorithm are universal and could be extended to other biomedical time series when none of the biomedical time series could be fully trusted.

It is worth diving deeper into the phase synchronization algorithm and carefully distinguishing it from the time-alignment approach, which is the most intuitive approach to align two signals. The distinction of time- and phase-alignment arises naturally from the fact that different EDR estimates come from the same breathing activity. To elaborate this fact, we model an EDR signal $f(t)$ by

$$f(t) = A(t)s\big(\phi(t)\big),\ \forall t\,, \tag{23}$$

where $s(t)$ is a 1-periodic function referred to as the wave-shape function [24]. In the above model, $A(t)$ is a smooth and positive function that controls the oscillatory strength of $f(t)$, $s(t)$ describes the oscillatory pattern of the EDR signal, and the phase function $\phi(t)$ is assumed to be strictly monotonically increasing and smooth, where $\phi'(t)$ is a positive function understood as the instantaneous frequency of $f(t)$. Usually we assume that $A(t)$ and $\phi'(t)$ vary slowly and the Fourier series coefficients of $s(t)$ decay fast, and we say that this signal fulfills the adaptive nonharmonic model (ANHM) [30]. Note that these assumptions are reasonable in the sense that the oscillatory pattern of a respiratory signal can usually be well captured by a few Fourier modes, and the amplitude and frequency do not vary fast. For more technical details on the model, we refer readers with interest to [30].

Given multiple EDR estimates $f_j, j = 1, \ldots, L$, we assume that they satisfy the form

$$f_j(t) = A_j(t) \cdot s_j\big(\phi(t) + \varphi_j/2\pi\big) + \epsilon_j(t), \qquad (24)$$

where $\epsilon_j(t)$ represents noise and $\varphi_j$ is the same phase deviation of the $j$-th EDR from the inaccessible breathing activity as the reference used in the above. Note that we could assume that $\epsilon_j(t)$ does not dominate the signal after the RQI-based selection. This comes from the fundamental fact that different EDR estimates come from the same breathing activity, so that all EDR estimates share the same phase up to a global phase deviation, which also says that the instantaneous frequency is the same for all EDR estimates. Note that each $\varphi_j$ is a shift as an element on the unit circle $S^1$. Specifically, if we write

$$f_j(t) = A_j(t) \cdot s_j\left(\phi(t) + \frac{\varphi_j}{2\pi}\right) = A_j(t) cos\big(2\pi\phi(t) + \varphi_j\big) + A_j(t) \sum_{l=2}^{\infty} cos\big(2\pi l\phi(t) + l\varphi_j\big) \qquad (25)$$

via the Fourier series expansion, we could assign the phase of $f_j(t)$ as $2\pi\phi(t) + \varphi_j$ [30].

To perform phase-alignment, we need to estimate $\varphi_j$ and undo this phase shift. To remove the shift, we count on the Hilbert transform $H$ to obtain its analytic representation. While it is well known that the Hilbert transform cannot fully recover the analytic representation of $f_j(t)$ [31] [32], the error is negligible when $f_j(t)$ fulfills the ANHM with a fast decay assumption of the Fourier series coefficients of $s_j$. With the fast decay assumption of the Fourier series coefficients of $s_j$ and the fact that we focus on the high quality EDR selected by RQI so that $\epsilon_j(t)$ does not dominate the signal, we obtain that $c_{i,j}$ in (16) satisfies $c_{i,j} \approx e^{i(\varphi_i - \varphi_j)}$. As a result, the synchronization gives $\tilde{\varphi}_j \approx \varphi_j + c$ for a global phase shift $c$ among all EDR estimates. As a result, the phase-shift of $f_j(t)$ can be undone by

$$\tilde{f}_j(t) = Re(e^{-\tilde{\varphi}_j} \cdot H\left(f_j(t)\right)). \qquad (26)$$

It is crucial to observe that the phase-alignment in (26) is fundamentally distinct from the intuitive time-shift approach. To appreciate the distinction, note that we could intuitively ask if we could find a constant $t_j$ for the $j$th EDR estimate, so that the time-shifted signal, denoted as $T_{t_j} f_j(t) := f_j\big(t - t_j\big) = A(t) \cdot s\big(\phi\big(t - t_j\big) + \varphi_j/2\pi\big) + \epsilon_j\big(t - t_j\big)$, could eliminate the phase discrepancy among different EDR estimates. Following the same idea of phase synchronization, we could view time shift as a sample in the Euclidean group $R$, estimate $t_j$, denoted by $\hat{t}_j$, by carrying out the time-synchronization via time-alignment, and then undo the time-shift. To this end, we can compute $\hat{t}_{ij}$ pairwise by minimizing the mean square error of $\left|f_j(t-t_{ij})-f_i(t)\right|$ over a preassigned range, like 5 seconds, which represents the typical period of respiration. Once all $\hat{t}_{ij}$ are computed, the $\hat{t}_j$ can be recovered using the same synchronization idea, which yield estimates of the form $f_j(t) := f_j(t-\hat{t}_j)$. However, this approach only works when $\phi$ is a linear function; that is, when the instantaneous frequency is fixed. When the instantaneous frequency $\phi'(t)$ is not constant, the instantaneous frequency will be shifted by $T_{t_j}$ as well, which leads to another discrepancy among different EDR estimates and downgrades the quality of the final ensemble. To illustrate this phenomenon and distinguish between phase- and time-alignment, see an illustration in Figure 3. In this example, six noisy phase-shifts of a signal with time-varying instantaneous frequency are simulated. While the phase-alignment can well recover a reference signal from each shifted signal, the time-alignment is limited. The difference can be clearly visualized in Figure 3(e) and (f), where we observe that disparities are especially apparent in the domain where instantaneous frequency increases. As a result, the final time-aligned ensemble fails to recover the desired reference signal.

We should mention that the fusion idea underlying our approach was previously considered in the literature. First, researchers have considered time-frequency analysis for the task of sensor fusion [33], where the spectrograms of different EDR estimates are averaged to obtain a robust estimate for the breathing rate. This method has merit for breathing rate estimation, but its potential for recovering the breathing pattern is not clear due to the lossy nature of spectrogram fusion. Fusion methods for respiratory rate estimate were investigated in [55], and results showed that the Smart Fusion algorithm [54] is the best one. The Smart Fusion approach enhances estimation over traditional fusion by removing signals with labeled artifact or low respiratory rate estimation quality. Note that a similar respiratory quality screening idea is utilized in our algorithm, where we count on the spectral information. Note that U4 in our sensitivity analysis catches the ideas of the Smart Fusion algorithm.

To validate our sync-ensembled EDR signal, we used three different signals (**CFLOW**, **THO**, **ABD**) as the ground truth in the CGMH database and **CFLOW** in the SHHS database. However, we note that none of these signals fully captures the inaccessible breathing activity, so the validation is only for the purpose of comparison amongst different EDR algorithms. A relevant interesting observation is that in Tables 1-4, while our proposed ensemble algorithm consistently outperforms others, the obtained $\gamma$-index and $\eta$-index are different when we compare with **CFLOW**, **THO** and **ABD.** This stems from the fact that the EDR signal, no matter which algorithm we use, usually has a wave-shape function (WSF) different from the WSF of **CFLOW**, **THO** or **ABD.**

The DM-based EDR algorithm deserves a discussion. It is a manifold learning approach that aims to recover the manifold that underlies the set of observed QRS complexes. The manifold model comes from the physiological fact that the QRS complexes are impacted by the breathing dynamics, and the impact can be mainly quantified by two parameters – the amplitude and width of QRS complexes. DM is designed to recover the eigenfunctions of the manifold [36], which allows us a recovery of the manifold [37], and hence the breathing dynamics. However, since DM is a nonlinear algorithm, one significant feature of using DM to extract breathing dynamics is that the breathing information can be spread nonlinearly across multiple eigenfunctions of the manifold. To be specific, it is not only the first non-trivial eigenfunction that contains the breathing information, but the other eigenfunctions do too. That is why we choose the top 5 eigenvectors as the DM-based EDR estimates. See [24] for an extensive discussion. This problem is also present when we attempt to use linear methods such as PCA to recover a nonlinear manifold. We furthermore note that DM is robust to additive noise near the manifold [27], which allows us a robust recovery of the top few eigenfunctions of the manifold. One issue that we do not address in this work is the linearity of the interpolation step. In principle, the interpolation should be done along the surface of the manifold. Since the focus of this work is ensembling multiple EDR estimates, algorithms for accomplishing this step will be explored in future work.

The ECG signal quality also deserves a discussion. Traditional EDR algorithms are dependent on accurate QRS detection results, which depends on the ECG signal quality and the chosen QRS detection algorithm. A false positive QRS detection results in a spike-like irregularity in the EDR signal. A false negative QRS detection typically results in the EDR signal becoming flat. An interesting alternative approach would be to build each individual EDR signal using a different QRS detection algorithm so that when ensembling takes place, these signal-specific irregularities would fail to propagate to the final EDR signal. In this work, only one QRS detection algorithm was used to create all EDR signals. It would be interesting to consider other R peak detection algorithms [40, 41] as other "soft sensors" of the respiratory signal and apply the fusion idea. EDR algorithms based on QRS slope, including DW and RA, also depend on the ECG signal quality. A recent comparison of EDR methods [56] demonstrated that methods derived from QRS slopes perform well compared to other EDR methods, but they are susceptible to the QRS complex morphology. Thus, they require clearly identified Q- and S-waves, together with a prominent R-wave peak [56]. In our implementation, the ECG signal quality was not screened, as the focus is our introduced ensembling algorithm.

Further studies to explore the practical value of sensor fusion in EDR are warranted. In particular, we should examine its application in remote monitoring or smartphone-enabled mobile healthcare. Nowadays, as sensor technologies advance, there are more and more easy-to-use, mobile-capable, and less-invasive sensors. While the qualities of these sensors tend to be questioned, the ensembling idea considered in this paper might be helpful to improve the overall quality of the collected signal. For example, it is well known that the photoplethysmogram (PPG) reflects not only hemodynamic status but also breathing activity [38]. Thus, we may consider combining breathing information extracted from an ECG signal with that extracted from a PPG signal [39] or even surveillance video to further improve the quality of the obtained respiratory signal. Conversely, the proposed sync-ensembled EDR may contain richer breathing information that could be better extracted with advanced signal processing tools; for example, we might apply time-frequency analysis [18] to extract instantaneous breathing rates or other respiratory quantities more accurately.

This work has several limitations. First, the proposed ensemble algorithm, and most other traditional EDR algorithms, mainly outputs a signal that is close to a sinusoidal one. Clearly this does not recover the morphology of any respiratory signal like **CFLOW, ABD** or **THO** (See Figures 3 and 4). Note that the oscillatory morphologies of different respiratory signals represent different features of the breathing dynamics. For example, the flow signal provides the inspiration and expiration times, and the end-tidal CO2 signal reflects the severity of patients with pulmonary disease. It is thus of great interest to have an algorithm that can recover the morphology of any respiratory signal of interest. To recover the morphology, we may consider a recent work [42] that recovers **CFLOW** from **ABD** and **THO** based on the Gaussian process and the amplitude and phase information of harmonics of **ABD** and **THO**. We leave this interesting topic as a future work. Since the kernel methods we employ (such as the diffusion maps algorithm) involve computing the top eigenvectors of $N \times N$ matrices (where $N$ is the number of heartbeats), the computational time required by our algorithm is high and increases as the length of the signal increases. For very long ECG signals where additionally non-stationarity will come into play, we suggest estimating the respiratory signal over short windows of 2 or 5 minutes. The recently proposed algorithm, Roseland (Robust and Scalable Embedding via LANmark Diffusion) [43], could be considered to speed up those diffusion-based algorithms. A second limitation of our study is that we do not explicitly handle premature ventricular contractions, which is a significant obstacle when estimating the amplitude modulation or QRS slope of the ECG signal. These morphologically distinct beats should be removed by an algorithm such as [44] before running any EDR algorithm. Our future work will include validating the proposed algorithm in a more general population.

## 6 CONCLUSION

We proposed and verified an ensembling approach based on the phase alignment via GCL and RQI to merge the information from different EDR algorithms. The resulting sync-ensembled EDR signal was demonstrated to be of superior quality. We envision the clinical value of our ensembling algorithm, particularly its application in digital health and remote medicine.

**Acknowledgement:** Jacob McErlean was supported by the U.S. National Science Foundation under Grant DGE 2139754. The Sleep Heart Health Study (SHHS) was supported by National Heart, Lung, and Blood Institute cooperative agreements U01HL53916 (University of California, Davis), U01HL53931 (New York University), U01HL53934 (University of Minnesota), U01HL53937 and U01HL64360 (Johns Hopkins University), U01HL53938 (University of Arizona), U01HL53940 (University of Washington), U01HL53941 (Boston University), and U01HL63463 (Case Western Reserve University). The National Sleep Research Resource was supported by the National Heart, Lung, and Blood Institute (R24 HL114473, 75N92019R002).

**Author contribution**

Jacob McErlean: idea, data collection, analysis and write-up
John Malik: idea, data collection, analysis and write-up
Yu-Ting Lin: idea, writing the manuscript.
Ronen Talman: idea, writing the manuscript.
Hau-Tieng Wu: idea, data collection, analysis and write-up

**Competing Interests statement**

No competing interests.

| EARR Scores | *T* | *PCA* | *DM* | *DW* | *RA* | *U* |
|---|---|---|---|---|---|---|
| **CFLOW** | 99.20 $\pm$ 1.96 | 98.92 $\pm$ 2.81 | 98.59 $\pm$ 3.03 | 98.98 $\pm$ 2.41 | 98.88 $\pm$ 2.70 | **99.14 $\pm$ 2.00** |
| **ABD** | 99.63 $\pm$ 0.77 | 99.61 $\pm$ 0.80 | 97.19 $\pm$ 1.51 | 99.53 $\pm$ 1.20 | 99.55 $\pm$ 1.05 | **99.76 $\pm$ 0.33** |
| **THO** | 97.24 $\pm$ 3.49 | 97.79 $\pm$ 2.73 | 97.56 $\pm$ 3.26 | 97.97 $\pm$ 2.77 | 97.51 $\pm$ 3.18 | 97.97 $\pm$ 3.46 |

| ɣ-Scores | *T* | *PCA* | *DM* | *DW* | *RA* | *U* |
|---|---|---|---|---|---|---|
| **CFLOW** | 85.31 $\pm$ 6.29 | 82.71 $\pm$ 7.06 | 80.99 $\pm$ 8.20 | 82.44 $\pm$ 8.36 | 83.96 $\pm$ 7.60 | **89.87 $\pm$ 4.79** |
| **ABD** | 77.86 $\pm$ 4.63 | 71.48 $\pm$ 5.74 | 68.05 $\pm$ 6.30 | 75.98 $\pm$ 5.82 | 76.47 $\pm$ 4.82 | **81.36 $\pm$ 3.71** |
| **THO** | 77.28 $\pm$ 7.81 | 72.01 $\pm$ 8.72 | 69.67 $\pm$ 8.93 | 74.72 $\pm$ 8.89 | 76.08 $\pm$ 7.89 | **81.85 $\pm$ 7.16** |

| OT Distances | *T* | *PCA* | *DM* | *DW* | *RA* | *U* |
|---|---|---|---|---|---|---|
| **CFLOW** | 0.55 $\pm$ 0.16 | 0.61 $\pm$ 0.18 | 0.62 $\pm$ 0.31 | 0.63 $\pm$ 0.26 | 0.58 $\pm$ 0.24 | **0.48 $\pm$ 0.11** |
| **ABD** | 0.73 $\pm$ 0.10 | 0.93 $\pm$ 0.13 | 0.95 $\pm$ 0.15 | 0.77 $\pm$ 0.16 | 0.70 $\pm$ 0.08 | **0.68 $\pm$ 0.07** |
| **THO** | 0.71 $\pm$ 0.18 | 0.84 $\pm$ 0.21 | 1.03 $\pm$ 0.24 | 0.78 $\pm$ 0.16 | 0.76 $\pm$ 0.21 | **0.62 $\pm$ 0.16** |

Table 1. Performance comparison on the CGMH database of the proposed sync-ensembled ECG-derived respiratory (EDR) signal and other methods evaluated by the estimated average respiratory rate (EARR) score, the γ-index, and the optimal transport (OT) distance. The EDR signals come from the traditional algorithm (T), principal component

analysis (PCA), diffusion map (DM), downward slope algorithm (DW), R-wave angle (RA) algorithm, and the sync-ensemble algorithm (U). Entries in the γ-index table (and EARR table) are the mean of γ-indices (and EARR scores) among the subjects ± the mean of standard deviations of γ-indices (and EARR scores) among the subjects. Entries in the OT distance table are the median of OT-indices among the subjects ± median absolute deviation of subject median absolute deviations of OT-indices among the subjects. For EARR scores, a statistically significant improvement, as determined by Wilcoxon signed rank tests at the 0.05 level with Bonferroni correction applied for multiple testing error, was observed by the sync-ensembled EDR over the individual signals, for the **CFLOW** signal; over T, DM, DW, and RA for **ABD** and **THO**. For γ-indices and OT distances, a statistically significant improvement was observed by the sync-ensembled EDR over the individual signals, for all reference signals **CFLOW**, **ABD**, and **THO**.

| **EARR Scores** | $U_1$ | $U_2$ | $U_3$ | $U_4$ | $U$ |
|---|---|---|---|---|---|
| **CFLOW** | 96.53 ± 7.68 | 99.10 ± 2.12 | **99.18 ± 1.81** | 99.16 ± 2.06 | 99.14 ± 2.00 |
| **ABD** | 97.88 ± 3.57 | 99.75 ± 0.33 | 99.73 ± 0.34 | 99.74 ± 0.33 | **99.76 ± 0.33** |
| **THO** | 94.99 ± 8.25 | 97.89 ± 3.07 | **98.43 ± 2.53** | 98.03 ± 3.52 | 97.97 ± 3.46 |

| **ɣ-Scores** | $U_1$ | $U_2$ | $U_3$ | $U_4$ | $U$ |
|---|---|---|---|---|---|
| **CFLOW** | 59.85 ± 1.14 | 86.46 ± 5.45 | 89.70 ± 4.78 | 86.59 ± 5.31 | **89.87 ± 2.00** |
| **ABD** | 52.48 ± 12.25 | 79.68 ± 4.19 | 81.06 ± 3.90 | 79.75 ± 4.05 | **81.36 ± 3.71** |
| **THO** | 54.22 ± 16.24 | 78.67 ± 7.25 | 81.58 ± 7.31 | 79.05 ± 7.18 | **81.85 ± 7.16** |

| **OT Distances** | $U_1$ | $U_2$ | $U_3$ | $U_4$ | $U$ |
|---|---|---|---|---|---|
| **CFLOW** | 2.83 ± 0.27 | 0.52 ± 0.11 | 0.48 ± 0.12 | 0.52 ± 0.11 | 0.48 ± 0.11 |
| **ABD** | 2.64 ± 0.33 | 0.68 ± 0.07 | 0.68 ± 0.08 | 0.69 ± 0.07 | 0.68 ± 0.07 |
| **THO** | 2.91 ± 0.30 | 0.70 ± 0.16 | 0.63 ± 0.17 | 0.70 ± 0.15 | **0.62 ± 0.16** |

Table 2. Sensitivity analysis on the CGMH database of the proposed sync-ensembled ECG-derived respiratory (EDR) signal and other methods evaluated by the estimated average respiratory rate (EARR) score, the $\gamma$-index, and optimal transport (OT) distance. The signal from the ensemble algorithm (U) is compared to four scenarios – the first a mean of EDR signal inputs ($U_1$); the second both without RQI selection and without phase-synchronization ($U_2$); the third without RQI selection but with phase-synchronization ($U_3$); and the fourth with RQI selection but without synchronization ($U_4$). Entries in the $\gamma$-index (and EARR) table are the mean of $\gamma$-indices (and EARR scores) among the subjects ± the mean of standard deviations of $\gamma$-indices (and EARR scores) among the subjects. Entries in the OT distance table are the median of OT-indices among the subjects ± median absolute deviation of subject median absolute deviations of OT-indices among the subjects. For all scoring metrics, a statistically significant improvement, as determined by Wilcoxon signed rank tests at the 0.05 level with Bonferroni correction applied for multiple testing error, was observed by U over $U_1$ for all reference signals **CFLOW**, **ABD**, and **THO**; by U over $U_2$ for the all reference signals for the $\gamma$-index and OT distance scores; by U over $U_3$ for **CFLOW** and **THO** for the $\gamma$-index, for **CFLOW** for the OT distance, and for **ABD** for the EARR score; and, by U over $U_4$ for all reference signals for the $\gamma$-index and OT distance.

| **EARR Scores** | *T* | *PCA* | *DM* | *DW* | *RA* | *U* |
|---|---|---|---|---|---|---|
| **CFLOW** | 97.76 ± 3.14 | 97.29 ± 3.45 | 94.79 ± 6.00 | 96.08 ± 5.40 | 96.52 ± 4.59 | **98.73 ± 1.86** |
| **ABD** | 97.41 ± 4.82 | 97.55 ± 4,82 | 94.45 ± 9.48 | 96.16 ± 7.06 | 96.47 ± 6.41 | **98.87 ± 2.14** |
| **THO** | 97.26 ± 5.09 | 97.22 ± 5.50 | 93.99 ± 9.80 | 96.12 ± 6.92 | 96.19 ± 6.93 | **98.84 ± 2.27** |

| **ɣ-Scores** | *T* | *PCA* | *DM* | *DW* | *RA* | *U* |
|---|---|---|---|---|---|---|
| **CFLOW** | 74.76±10.32 | 73.05 ± 10.32 | 68.43 ± 11.33 | 68.88 ± 10.62 | 70.67 ± 10.11 | **84.13 ± 7.58** |
| **ABD** | 78.41±14.90 | 78.12 ±14.99 | 72.27 ± 21.08 | 73.26 ± 17.98 | 75.60 ± 16.90 | **88.97 ± 8.65** |
| **THO** | 76.80±15.33 | 76.26 ± 15.83 | 70.29 ± 21.25 | 71.55 ± 18.49 | 73.92 ± 17.67 | **87.79 ± 9.45** |

| **OT Distances** | *T* | *PCA* | *DM* | *DW* | *RA* | *U* |
|---|---|---|---|---|---|---|
| **CFLOW** | 1.73 ± 0.23 | 1.82 ± 0.26 | 1.88 ± 0.28 | 1.92 ± 0.22 | 1.85 ± 0.24 | **1.41 ± 0.23** |
| **ABD** | 0.91 ± 0.32 | 0.90 ± 0.32 | 0.95 ± 0.38 | 1.05 ± 0.39 | 0.93 ± 0.34 | **0.66 ± 0.19** |

| THO | 1.04 ± 0.32 | 1.00 ± 0.33 | 1.12 ± 0.40 | 1.24 ± 0.36 | 1.09 ± 0.34 | **0.74 ± 0.22** |
|---|---|---|---|---|---|---|

Table 3. Performance comparison on the SHHS database of the proposed sync-ensembled ECG-derived respiratory (EDR) signal and other methods evaluated by the estimated average respiratory rate (EARR) score, the γ-index, and the optimal transport (OT) distance. The EDR signals come from the traditional algorithm (T), principal component analysis (PCA), diffusion map (DM), downward slope algorithm (DW), R-wave angle (RA) algorithm, and the sync-ensemble algorithm (U). Entries in the γ-index table (and EARR table) are the mean of γ-indices (and EARR scores) among the subjects ± the mean of standard deviations of γ-indices (and EARR scores) among the subjects. Entries in the OT distance table are the median of OT-indices among the subjects ± median absolute deviation of subject median absolute deviations of OT-indices among the subjects. For all scoring metrics, a statistically significant improvement, as determined by Wilcoxon signed rank tests at the 0.05 level with Bonferroni correction applied for multiple testing error, was observed by the sync-ensembled EDR over the individual signals, for all reference signals **CFLOW, ABD,** and **THO**.

| **EARR Scores** | $U_1$ | $U_2$ | $U_3$ | $U_4$ | $U$ |
|---|---|---|---|---|---|
| **CFLOW** | 94.26 ± 8.84 | 97.70 ± 3.37 | 98.37 ± 2.13 | 98.58 ± 2.08 | **98.73 ± 1.86** |
| **ABD** | 94.40 ± 10.76 | 97.67 ± 4.50 | 98.14 ± 3.72 | 98.76 ± 2.33 | **98.87 ± 2.14** |
| **THO** | 94.19 ± 10.97 | 97.51 ± 5.03 | 98.03 ± 4.19 | 98.72 ± 2.54 | **98.84 ± 2.27** |

| **ɣ-Scores** | $U_1$ | $U_2$ | $U_3$ | $U_4$ | $U$ |
|---|---|---|---|---|---|
| **CFLOW** | 53.49 ± 15.78 | 78.80 ± 8.81 | 82.33 ± 8.27 | 81.71 ± 7.87 | **84.13 ± 7.58** |
| **ABD** | 56.90 ± 18.71 | 83.11 ± 12.35 | 87.03 ± 10.78 | 86.07 ± 9.32 | **88.97 ± 8.65** |
| **THO** | 56.18 ± 18.81 | 81.54 ± 13.51 | 85.36 ± 12.10 | 85.07 ± 10.01 | **87.79 ± 9.45** |

| **OT Distances** | $U_1$ | $U_2$ | $U_3$ | $U_4$ | $U$ |
|---|---|---|---|---|---|
| **CFLOW** | 3.26 ± 0.21 | 1.53 ± 0.25 | 1.49 ± 0.24 | 1.49 ± 0.25 | **1.41 ± 0.23** |
| **ABD** | $2.99 \pm 0.30$ | $0.75 \pm 0.28$ | $0.69 \pm 0.25$ | $0.70 \pm 0.20$ | $\mathbf{0.66 \pm 0.19}$ |
| **THO** | $2.94 \pm 0.27$ | $0.86 \pm 0.29$ | $0.77 \pm 0.27$ | $0.78 \pm 0.23$ | $\mathbf{0.74 \pm 0.22}$ |

Table 4. Sensitivity analysis on the SHHS database of the proposed sync-ensembled ECG-derived respiratory (EDR) signal and other methods evaluated by the estimated average respiratory rate (EARR) score, the γ-index, and optimal transport (OT) distance. The signal from the ensemble algorithm (U) is compared to four scenarios –he first a mean of EDR signal inputs ($U_1$); the second both without RQI selection and without phase-synchronization ($U_2$); the third without RQI selection but with phase-synchronization ($U_3$); and the fourth with RQI selection but without synchronization ($U_4$). Entries in the γ-index (and EARR) table are the mean of γ-indices (and EARR scores) among the subjects ± the mean of standard deviations of γ-indices (and EARR scores) among the subjects. Entries in the OT distance table are the median of OT-indices among the subjects ± median absolute deviation of subject median absolute deviations of OT-indices among the subjects. For all scoring metrics, a statistically significant improvement, as determined by Wilcoxon signed rank tests at the 0.05 level with Bonferroni correction applied for multiple testing error, was observed by U over $U_1$, $U_3$, $U_3$, and $U_4$, for all reference signals **CFLOW, ABD,** and **THO**.

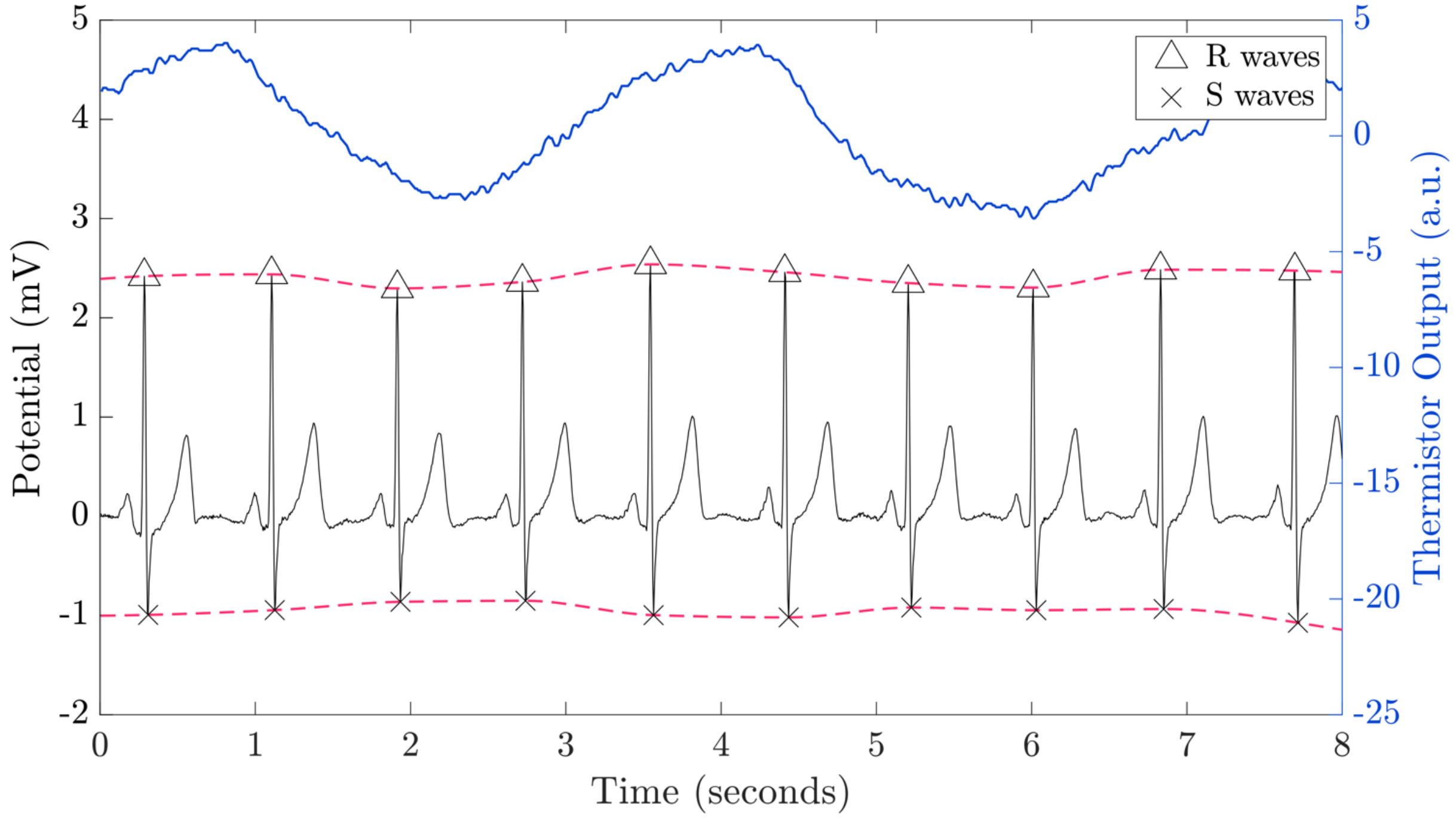

Fig. 1. An indirect respiratory signal can be derived from an ECG recording by measuring the amplitudes of each ventricular depolarization (QRS) complex. This amplitude is traditionally recorded as the difference in potential between the R wave and the S wave. Top: a ground truth respiratory signal (nasal thermistor recording temperature differential) in blue. Bottom: the simultaneously recorded ECG signal and its amplitude modulation in the dotted pink lines that gives the traditional EDR.

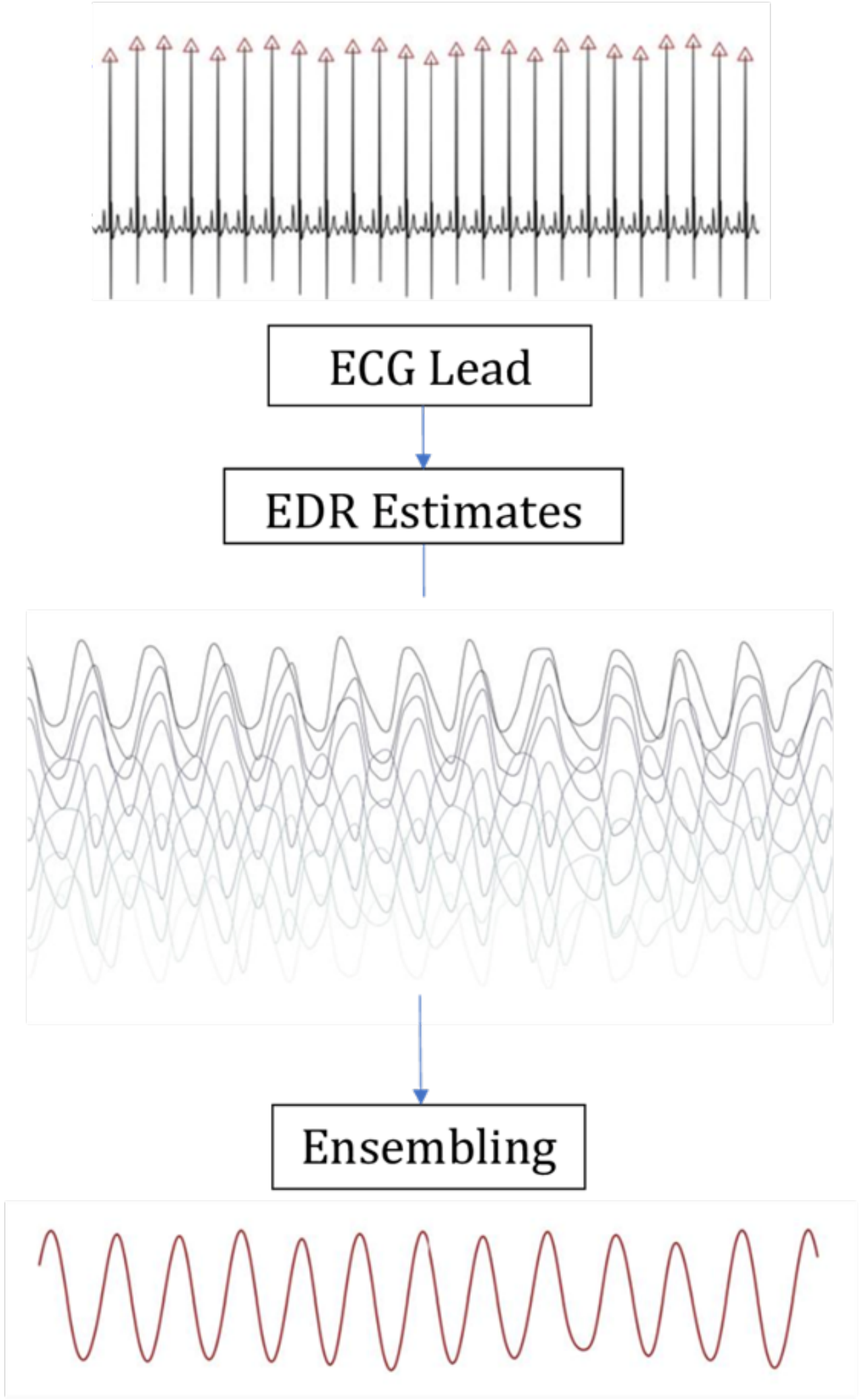

Fig. 2. The main steps in the proposed sync-ensembled EDR algorithm, in which multiple EDR signals (estimated using diverse methods) are fused to form a stronger one. In the middle, we show the individual EDR estimates which, while different, share the information that can be extracted by the ensembling method. All high quality EDR estimates are then ensembled to obtain the final EDR signal. We note that this method can in theory be extended when more ECG leads are available.

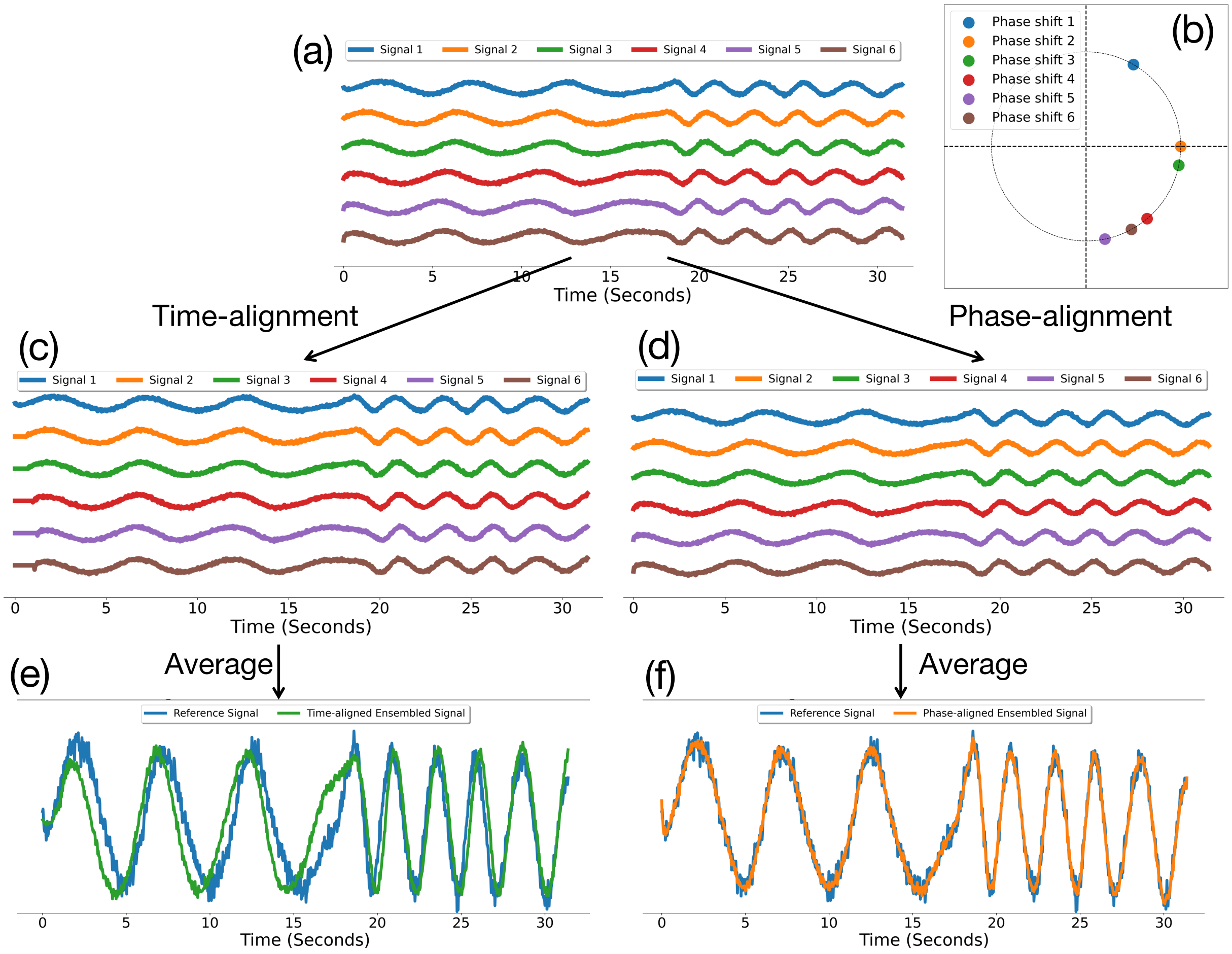

Fig. 3. A comparison of time- and phase- alignments. (a) six signals with different global phase shifts, where the phase shifts for the shifted signals (in radians) are approximately 1.05, 0.00, -0.20, -0.8, -1.37, and -1.07 as are shown in (b). (c) and (d) are the results of time- and phase- alignment, and (e) and (f) are the results of taking average of the time- and phase- aligned signals, where the true signal before phase shift is superimposed for a visual comparison.

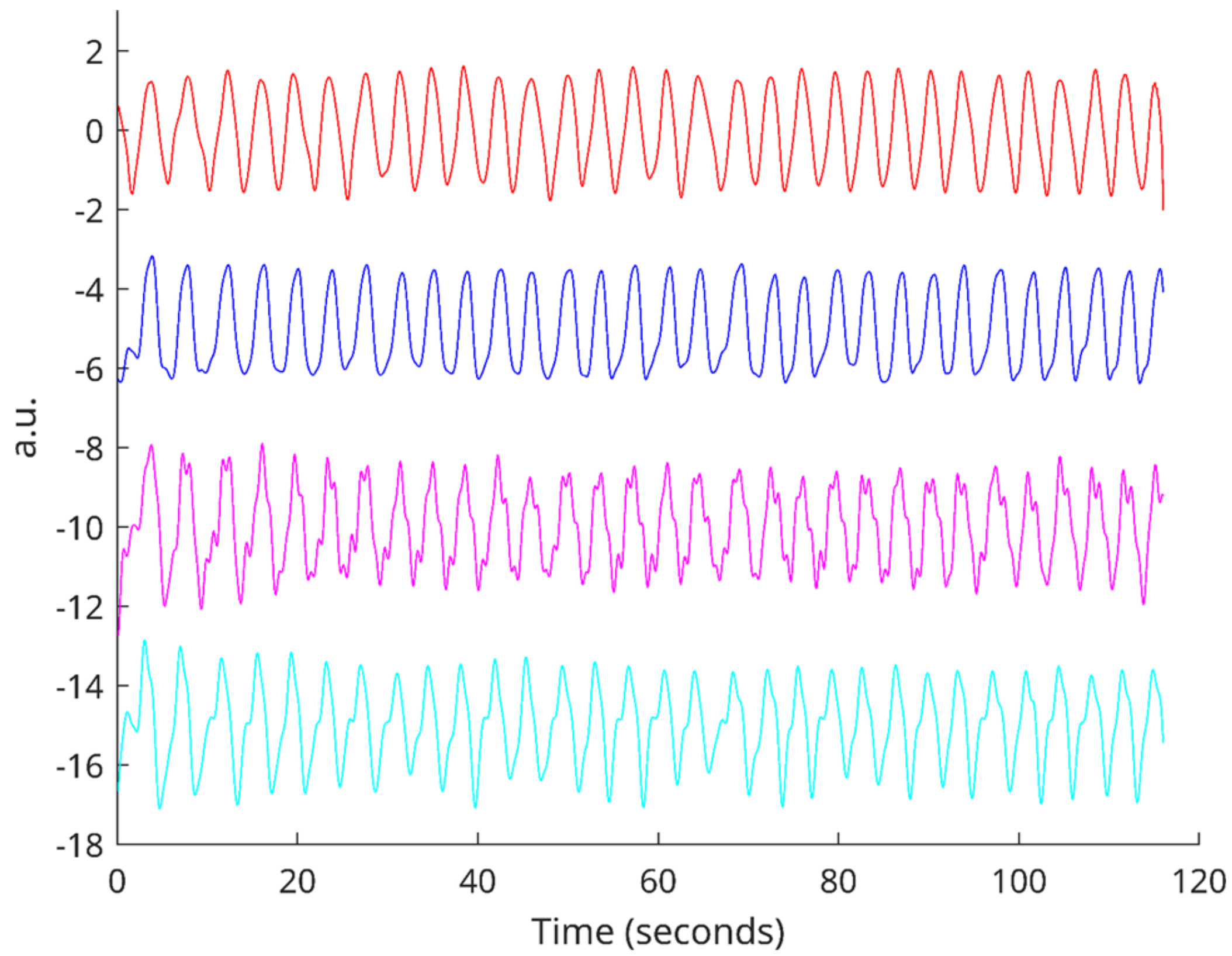

Fig. 4. A comparison of the EDR derived by the proposed sync-ensemble EDR algorithm and other respiratory signals. It shows that the proposed EDR could accurately measure the respiratory cycle. From top to bottom, we show: the sync-ensembled EDR signal; the THO signal; the ABD signal; and the CFLOW signal.

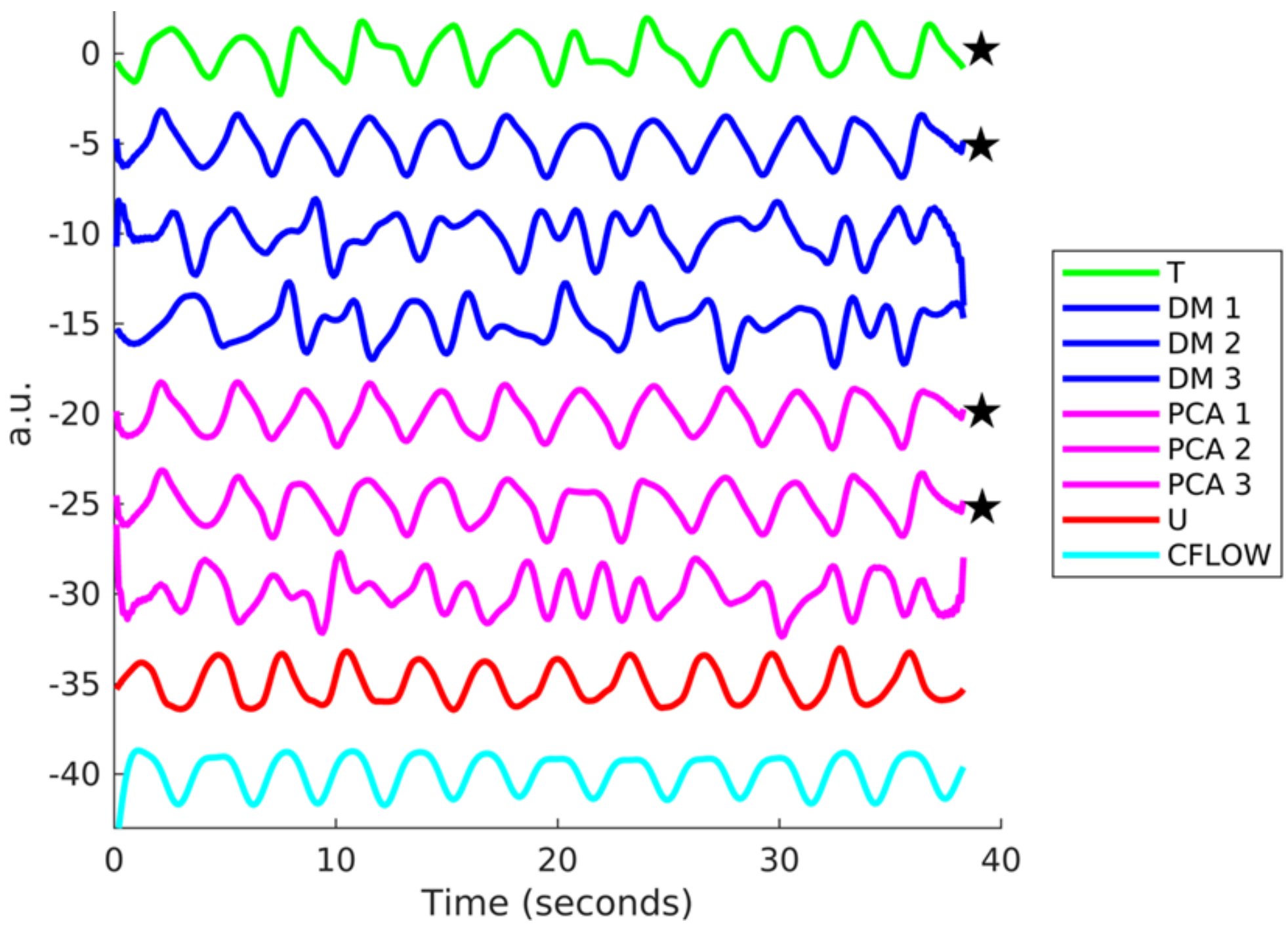

Fig. 5. A comparison of EDR signals derived by different algorithms, including the proposed sync-ensemble EDR algorithm. From top to bottom, we show: the traditional EDR signal (T), the top 3 DM-based EDR signal; the top 3 PCA-based EDR signal, the sync-ensembled EDR signal (U), and the CFLOW signal. Those EDR signals marked by star are the selected high quality EDR signals in the sync-ensemble EDR algorithm. Visually, the traditional EDR signal is worse than other chosen EDR signals and the final result U.

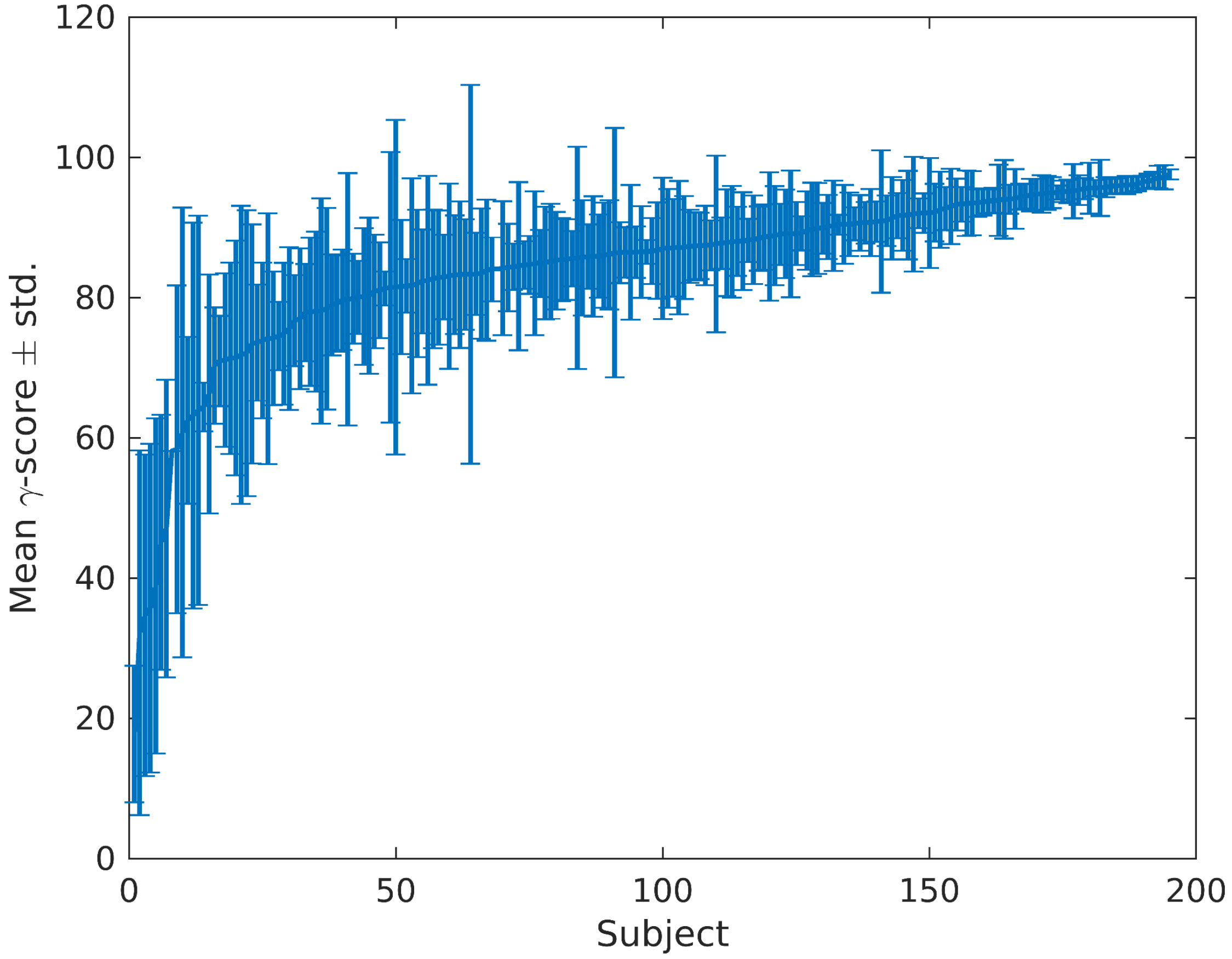

Fig. 6. An errorbar plot of the mean $\gamma$-score $\pm$ standard deviation for each of the subjects from the SHHS database, ordered by the increasing mean $\gamma$-score.

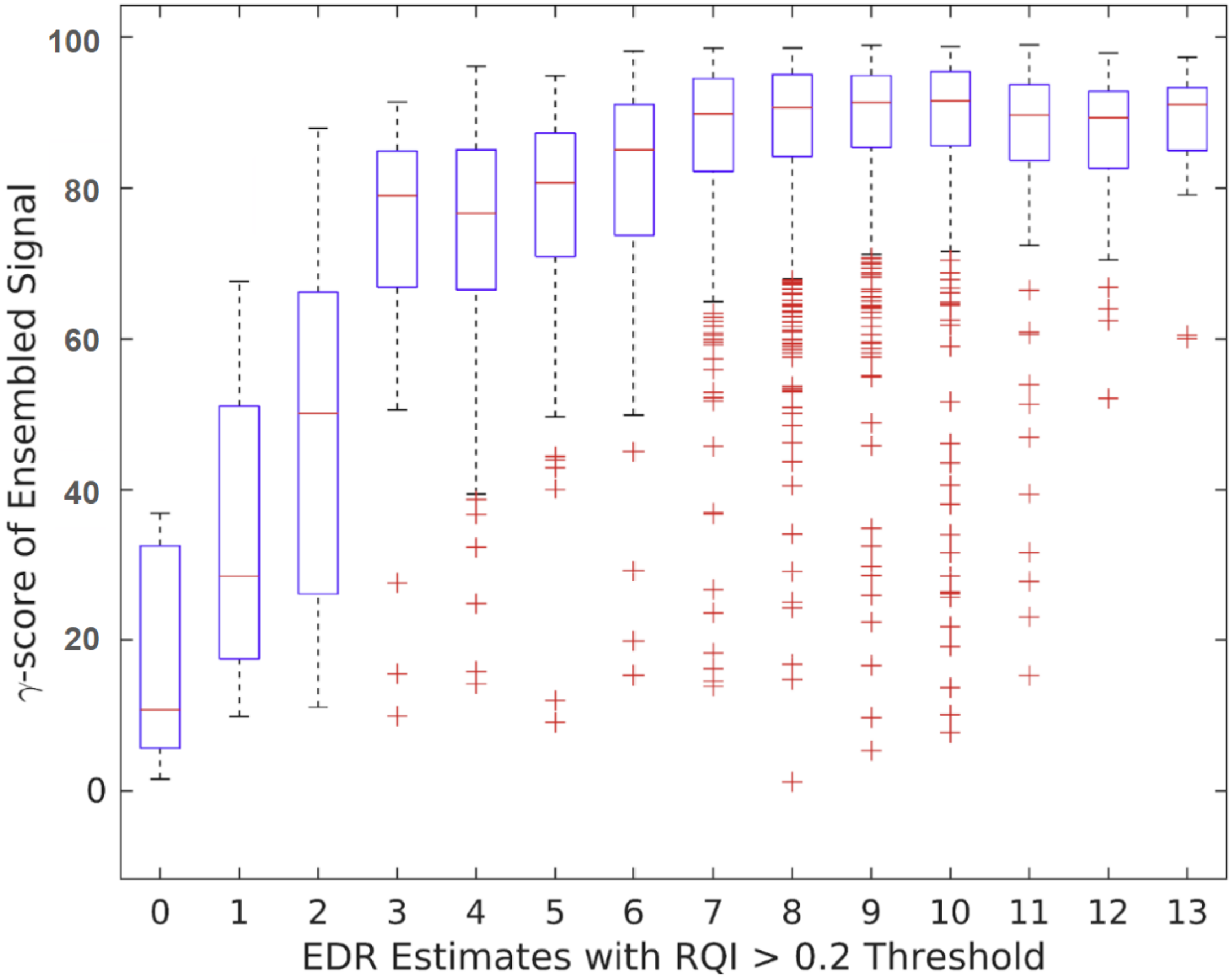

Fig. 7. For all (3220) 2-minute CFLOW segments from the SHHS database, we compare the number (out of 13) of selected EDR signals above the RQI threshold of 0.2 to the $\gamma$-score of the resulting ensembled signal.

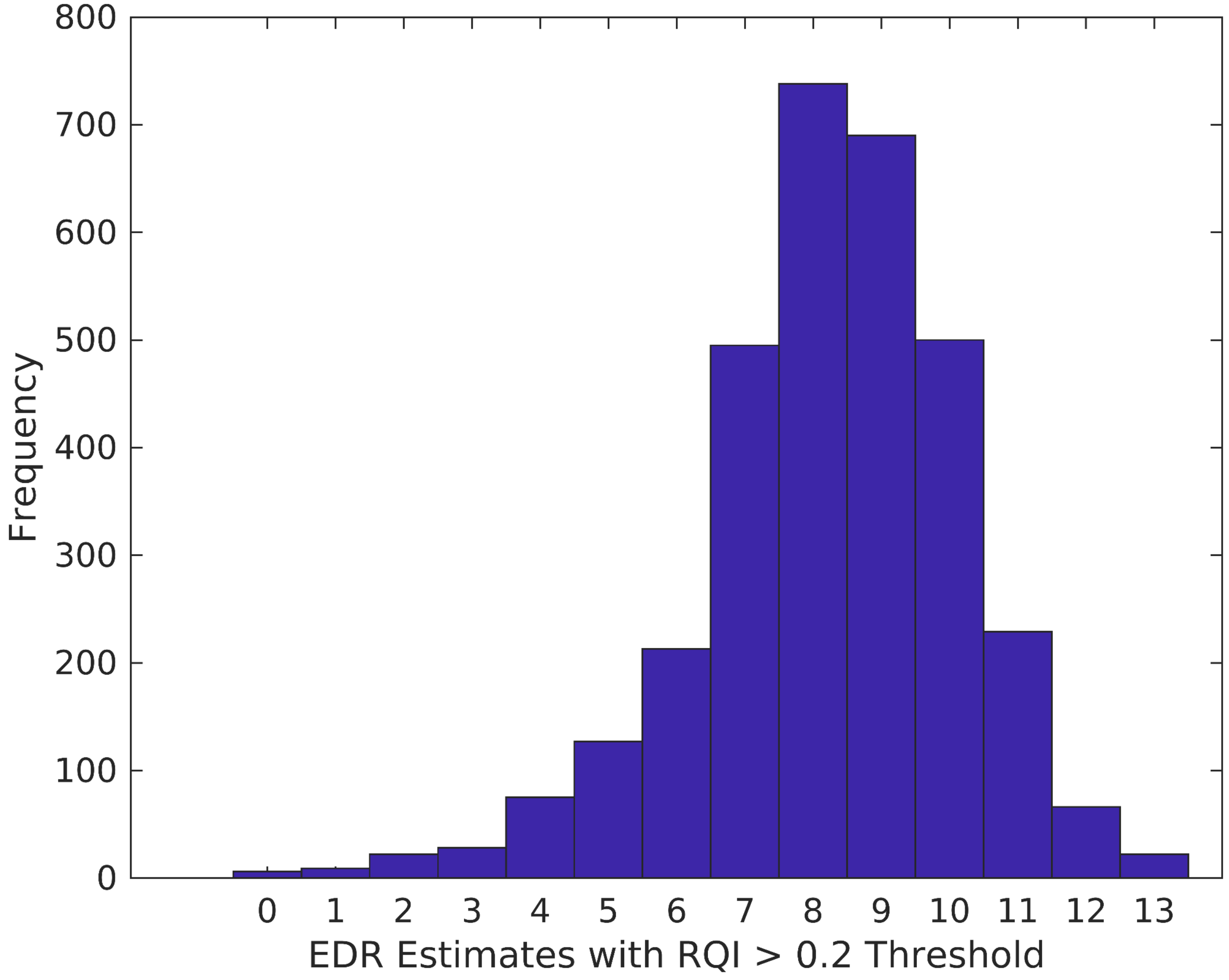

Fig. 8. For all 2-minute CFLOW segments from the SHHS database, the distribution of the number (out of 13) of EDR signals which are above the 0.2 RQI threshold.